\documentstyle[aasms4,12pt]{article}
\begin{document}

\title{A MULTIPLE MERGER MODEL FOR THE ORIGIN OF ULTRALUMINOUS INFRARED GALAXIES} 

\author{Yoshiaki Taniguchi, and Yasuhiro Shioya}

\vspace {1cm}

\affil{Astronomical Institute, Tohoku University, Aoba, Sendai 980-8578, Japan}


\begin{abstract}

It has been often considered that the dissipative collapse caused by a merger
of two gas-rich galaxies is responsible for the intense nuclear starbursts or the 
nonthermal quasar activity in ultraluminous infrared galaxies due to the efficient
fueling induced by it. It is also widely known that Ultraluminous Infrared Galaxies
(ULIGs) are often found in merging systems. Some ULIGs, such as Arp 220, show two 
compact starburst regions which are considered to be associated with two galactic 
nuclei in the process of merging. However, since a merger between two galaxies may
make only one compact starburst region, we suggest the possibility that 
double-nucleus ULIGs are composed of two merging nuclei, each of which contains a 
couple of galactic  nuclei.

\end{abstract}




\section{INTRODUCTION}

Ultraluminous Infrared Galaxies (hereafter ULIGs) have attracted much attention since 
their discovery by {\it IRAS} in 1984 (Soifer et al. 1984; Wright, Joseph, \& Meikle 1984;
see for a review Sanders \& Mirabel 1996). 
The multiple morphological studies published to date have shown that these objects tend to
be found in galaxy mergers or in strongly interacting galaxies 
(Sanders et al. 1988a; Lawrence et al. 1991; Leech et al. 1994).
Although the origin of their huge infrared luminosities is still
not fully understood, it is considered to come from intense starbursts, central active
galactic nuclei or a combination of both 
(Joseph \& Wright 1985; Sanders et al. 1988a, 1988b; Solomon \& Sage 1988;
Scoville et al. 1991; Condon et al. 1991; Majewski et al. 1993; Lonsdale et al.
1994; Skinner et al. 1997).
As numerical simulations have shown (see for a review Shlosman, Begelman, \& 
Frank 1990; Barnes \& Hernquist 1992), galaxy mergers cause efficient gas fueling 
toward the nuclear regions 
of the merging systems that ultimately can trigger and maintain any of the central 
activities mentioned above, either as a result of the piling of gas 
(Negroponte \& White 1983; Barnes 1988; Barnes \& Hernquist 1991;
Olson \& Kwan 1990a, 1990b;
Noguchi 1991; Bekki \& Noguchi 1994; Mihos \& Hernquist 1994a, 1994b; Hernquist \& Mihos
1995) or by the dynamical effect of supermassive binaries
(Taniguchi \& Wada 1996; Taniguchi 1997; Taniguchi, Wada, \& Murayama 1997).

Current models for the origin of ULIGs only consider the merging of {\it two} 
gas-rich galaxies (Sanders et al. 1988a; Kormendy \& Sanders 1992). However, taking into
account that there are a large number of nearby compact galaxy groups (e.g., Hickson 
1982), formation of ULIGs due to a multiple merger cannot be ruled out. 
In fact, the presence of three OH maser components in the archetypical ULIG Arp 220 suggests
the possibility that this ULIG originates from a multiple merger (Diamond et al. 1989).
In this {\it Letter}, we discuss the multiple merger scenario for the formation of ULIGs
based on the observational properties of Arp 220.

\section{THE COMPACT STARBURSTS IN ARP 220}

Arp 220, being the nearest double-nucleus ULIG known, has been subject to observations
in a wide range of wavelengths during the last decade. A brief summary of the observational
properties of the nuclear starbursts in this object\footnote{We use
a distance of 74 Mpc (1 arcsec = 359 pc) derived from its recession velocity
to the Galactic Standard of Rest of $V_{\rm GSR}$ = 5,531 km s$^{-1}$ (de Vaucouleurs et al.
1991) and a Hubble constant of $H_0$ = 75 km s$^{-1}$ Mpc$^{-1}$.} is given below.

Radio continuum images show the presence of two compact starburst regions (sizes $\leq$ 100
pc)(Norris 1988; Condon et al. 1991) located in the nuclear region of this object with a 
projected separation of 0.95 arcsec (350 pc). They are surrounded by a circumnuclear 
post-starburst region with a spatial extent of 1.4 kpc $\times$ 0.7 kpc
in which stars with ages of $\sim 10^8$ years are dominant (Shaya et al. 1994; Larkin et al.
1995; Armus et al. 1995).Taking into account that typical starburst nuclei
have ionized regions with sizes of $\sim$ 500 pc to 1 kpc (e.g., 
Sugai \& Taniguchi 1992), it is clear that the nuclear starbursts in this object are
quite outstanding in the sense that they are able to produce huge luminosities from quite 
more compact regions ($\simeq$ 100 pc).

It has been known for a while that starburst activity is associated to high-density gas 
components (traced by high-density tracer molecules like HCN or HCO$^+$). We therefore expect
that the nuclear starbursts in this object also are related to dense gaseous
systems formed during the course of the merger that created it. In order to demonstrate the 
presence of high-density gas around the two starburst regions in Arp 220, we show the profiles
of the HCN and HCO$^+$ emission lines (Solomon, Downes, \& Radford 1992)
in Figure 1. Although the signal-to-noise ratios of these observations are not so high, both 
clearly show double-peaked profiles. The line peaks lie at 5300 km s$^{-1}$ and 5600 km s$^{-1}$,
which roughly correspond to the recession velocities of the eastern and the western nuclei, 
respectively (Larkin et al. 1995), as shown in the bottom panels of Figure 1. These two peaks
can also be seen at similar velocities in the CO($J$=2-1) map by Scoville, Yun, \& Bryant 
(1997). Although there seems to be a velocity difference in the western component between the 
molecular gas and Pa$\alpha$ emission (5400 km s$^{-1}$ in Larkin et al. 1995), it should be 
noted that the large extinction toward the central region of Arp 220 may be also affecting the
latter and that a velocity difference of $\sim$ 100 km s$^{-1}$ does not affect qualitatively
our discussion below. We therefore conclude that the two compact starbursts occurred in the 
dense gas media surrounding the two nuclei.

The fact that these nuclear starbursts are associated to high-density regions can naturally 
explain their compactness. As is widely known, given the ionizing flux and the local density
of a region, the radius of the equivalent fully ionized zone can be derived (e.g., Osterbrock
1989). The number of ionizing photons in the nuclear region of Arp 220 is estimated from the 
de-reddened Br$\gamma$ emission-line flux as $2\times 10^{-14}$ erg s$^{-1}$ cm$^{-2}$ (Shier 
et al. 1994). Since on average 77 ionizing photons produce one Br$\gamma$ photon, we obtain an
ionizing photon rate of $Q({\rm H}^0) \simeq 1.1 \times 10^{54}$ photons s$^{-1}$. Therefore,
the radius of the ionized nebula will be $r_{\rm neb} = [(3/4\pi) n_{\rm H}^{-2} 
\alpha_{\rm B}^{-1}  Q({\rm H}^0)]^{1/3} \sim 50 (n_{\rm H}/ 500~ {\rm cm}^{-3})^{-2/3}$ pc, 
where $n_{\rm H}$ is the number density of hydrogen in the nebula and $\alpha_{\rm B}$ is the
effective recombination coefficient for hydrogen ($\alpha_{\rm B} = 2.59\times 10^{-13}$ cm$^3$
s$^{-1}$ for an electron temperature of $T_{\rm e} = 10^4$ K, Osterbrock 1989).
Since ULIGs tend to have larger amounts of high-density gas (e.g., $10^{4{\rm -}5}$ cm$^{-3}$) 
than normal galaxies (Solomon, Downes, \& Radford 1992), it is expected that their average 
molecular cloud gas density in their nuclear regions will exceed $\sim 10^2$ cm$^{-3}$ 
significantly. In fact, Scoville et al. (1997) estimated an average molecular hydrogen density
of $n_{\rm H_2} \simeq 2\times 10^4$ cm$^{-3}$ in the central 130 pc region of Arp 220. The 
filling factor-corrected number density of hydrogen of this high-density component can be 
estimated from the fact that typical starburst nuclei have a few $10^8 M_\odot$ of molecular
gas within their central $\sim$1 kpc regions (Devereux et al. 1994), giving an average hydrogen
number density of at most $n_{\rm H} \sim$ 50 cm$^{-3}$ (an order of magnitude larger than the
large scale average for a typical disk of a normal galaxy) and therefore a radius for the
ionized nebulae of $r_{\rm neb} \simeq 230$ pc.
We may thus conclude that the compactness of the nuclear starbursts in the ULIGs
may be due to the higher gas density in their nuclear regions.

\section{DISCUSSION}

\subsection{Origin of the Nuclear Starbursts in ULIGs}

Numerical simulations of galaxy mergers between two gas-rich disk galaxies have shown that they
induce efficient gas fueling into the central a few 100 pc region of the
merging systems (e.g., Mihos \&  Hernquist 1994b and references therein). However, it is known 
that the properties of the merger-driven starbursts are sensitive to the structure of the 
progenitor galaxies (Mihos \& Hernquist 1994b). If the progenitor galaxies are bulgeless (i.e.,
late-type spirals), the successive close encounters during the merger strongly affect their 
respective gas disks. As the merger proceeds, gas clouds in each galaxy are channeled to each 
nuclear region and  if a dynamical instability occurs in the central region of each member, 
intense star formation would occur there (e.g., Noguchi 1991; Shlosman \& Noguchi 1994). Since
the gas contained in the progenitors would be used up while the merger is in progress, no 
intense starbursts would occur when the merger is complete. The well-known merger, NGC 7252,
may be a good example for this case because only moderate star-forming regions can be seen
in the central part of this galaxy (Whitmore et al. 1993). On the other hand, if the galactic
bulge of each progenitor is massive enough to stabilize their nuclear gas disks, it will 
prevent bulge strong gas inflows until the galaxies merge, giving rise to a single 
intense starburst in the central part of the merger remnant (see also Bekki \& Noguchi 1993).
Since in the case of Arp 220, the projected separation between the
two compact starburst regions is 350 pc, it is suggested that at least if  double-nucleus
ULIGs come from mergers between two galaxies, their progenitors could be gas-rich galaxies 
without prominent bulges.

The double-nucleus (traced by radio continuum and NIR emission) nature of Arp 220 has often 
used as an example of this two-galaxy merger scenario 
(Baan \& Haschick 1987; Norris 1988; Sanders et al. 1988a; Graham et al. 1990;
Scoville et al. 1998)
However, there is some observational evidences that Arp 220 may come in 
fact from the merger of more than two spirals. VLBI mapping of OH megamaser emission in Arp 
220 has revealed the presence of at least three bright OH megamaser spots in its nuclear region
(Diamond et al. 1989), which would imply the coexistence of
three active galactic nuclei in the region. Therefore, given the fact that there exist
a fairly large amount of nearby compact galaxy groups that could end up merging since their 
merging timescales are generally shorter than the Hubble time 
(Barnes 1989; Weil \& Hernquist 1996),
we cannot rule out the multiple merger scenario for the formation of ULIGs.

For simplicity, we consider the case of ULIG coming from a merger of four comparably 
nucleated disk galaxies. At an early stage of the merger it is expected that parings would
occur; i.e., two pairs of galaxies merge first. Then the merging remnants would merge again
into one final object (cf. Barnes 1989; Weil \& Hernquist 1996).
It is known that a merger between two {\it nucleated} galaxies can form a
single dense gaseous system in the merger remnant because of the dynamical 
disturbance of the binary potential to the gas clouds  (Bekki \& Noguchi 1994; Taniguchi \& 
Wada 1996). If double-nucleus ULIGs are on the way to the final merger,
we can explain why they have two dense gaseous systems each of which
can be associated with a merger remnant between two galaxy nuclei.

\subsection{A Multiple Merger Model for Arp 220}

We now consider more carefully the possibility that Arp 220 comes from a merger
of four comparably nucleated disk galaxies. In this scenario, the double nucleus of this
object should correspond to the final stages of the merging of two pairs of nuclei. As 
mentioned above, there are at least three bright OH megamaser spots
in the nuclear region of Arp 220 (Diamond et al. 1989). The eastern nucleus contains two
bright OH maser spots with a projected separation of 47.6 pc while the
western shows only one component. Recent VLBI measurements by Lonsdale et al. (1994) 
have shown that the western (i.e., the brightest) component of the OH maser
originates from a very compact region whose size is less than 1 pc.
This measurement suggests strongly that the western component is produced by the pumping
by far-infrared continuum emitted by a dusty torus around an active galactic nucleus rather
than by the luminous nuclear starbursts. If this is also the case for the two eastern OH
maser components, Arp 220 would contain at least three active galactic nuclei (i.e., three
supermassive black holes) as suggested by Diamond et al. (1989). It is of course possible
that the single western OH megamaser component actually represents the accidental alignment
along our line of sight of two nuclei, which would be resolved if observed in other
conditions.

Another support for the multiple merger scenario comes from the observational fact that the
two OH maser spots in the eastern nucleus show a velocity gradient which is almost 
perpendicular to the eastern-western nucleus axis, implying that the eastern nucleus
is dynamically different from the western nucleus (Diamond et al. 1989).
Furthermore, H$_2$CO maser observations of this object (Baan \& Haschick 1995) show
velocity gradients in maser emission associated with both the eastern and western nuclei
that are more compatible with rotational motion around the individual nuclei
rather than with the global rotation around the two nuclei (see Baan \& Haschick 1995).

In our scenario it should also be expected that each pair of nuclei has a rotating 
gas disk settled roughly in the orbital plane of each of the two nuclei in each pair
because nuclear gas will settle in a relatively short timescale there. In fact, the
double-peaked nature of the HCN and HCO$^+$ (Solomon et al. 1992) and CO($J$=1-0) emissions
(Scoville et al. 1997) suggest that the dense gaseous systems are associated with each
pair of merging nuclei. At larger scales, a circumnuclear gas disk with a radius of 
$\sim$ 300 pc surrounds the two pairs of nuclei (Scoville et al. 1997).

In summary, the multiple merger scenario that we proposed here explains almost all the
observational properties of Arp 220 consistently. A schematic illustration of the 
nuclear region of Arp 220 is shown in Fig. 2. The rotation of the eastern black hole
binary was determined from the velocity difference between the two OH megamaser components,
IIa and IIb (Diamond et al. 1989) while that of the eastern one is from the rotation of the
H$_2$CO masing gas (Baan \& Haschick 1995). The global rotation of the E and W nuclei is
from the NIR spectroscopy (Larkin et al. 1995). Although any current observational 
facilities may not be able to verify this model, we hope that the multiple merger 
scenario will be taken into account in the future study on the origin of ULIGs.

\vspace{0.5cm}

We would like to thank Dave Sanders, Baltasar Vila-Vilaro, Sumio Ishizuki, Seiichi Sakamoto,
and Neil Trentham for useful discussion and suggestions.
We also thank John Hibbard and Dave Sanders for
providing us with their CCD image of Arp 220. This work was financially supported in 
part by Grant-in-Aids for the Scientific Research (No. 0704405) of the Japanese Ministry
of Education, Culture, Sports and Science.


\newpage


\centerline {\bf Figure captions}\par

\vspace{1cm}

\noindent {\bf Fig. 1:} $^{12}$CO($J$=1-0), HCN, and HCO$^+$ emission profiles
taken from Solomon et al. (1992) are shown in the top, the second and the third
panel, respectively. The rotation curve of the nuclear region (Larkin et al. 1995)
and the double radio components which show the two compact starburst regions
are also shown in the bottom left and right panels, respectively. The double 
peaks of the HCN, and HCO$^+$ emission appear to correspond to the two
starburst regions kinematically.

\vspace{1cm}

\noindent {\bf Fig. 2:} A schematic picture of the multiple merger model
of Arp 220. The direct $R$-band CCD image shown in the left panel 
was kindly supplied by J. Hibbard and D. Sanders. The two compact starbursts
traced by the the radio continuum imaging at 8.44 GHz (Condon et al. 1991) are shown
in the lower middle panel. The two right panels show the tangential and the
line-of-sight view of Arp 220 based on the multiple merger model.

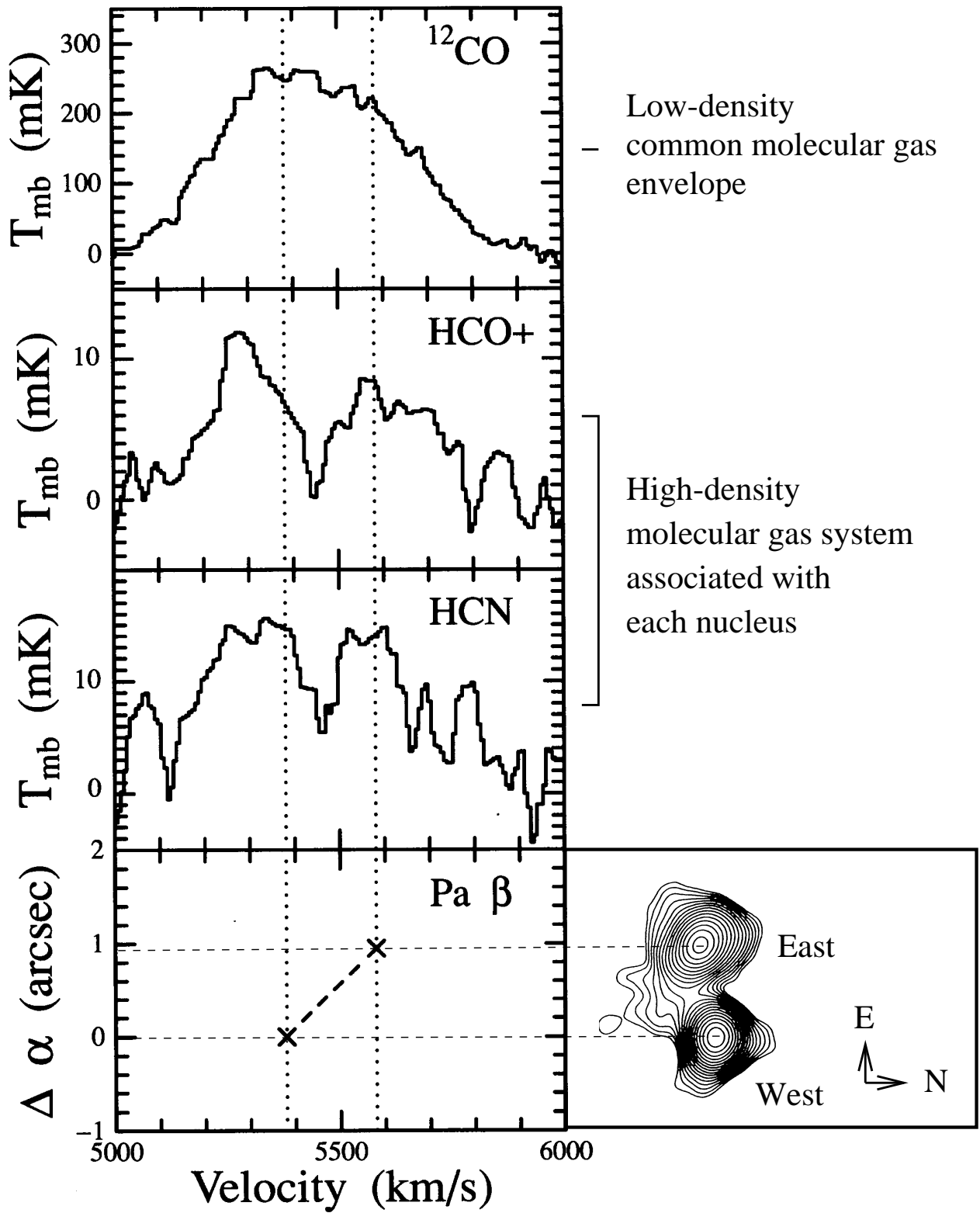
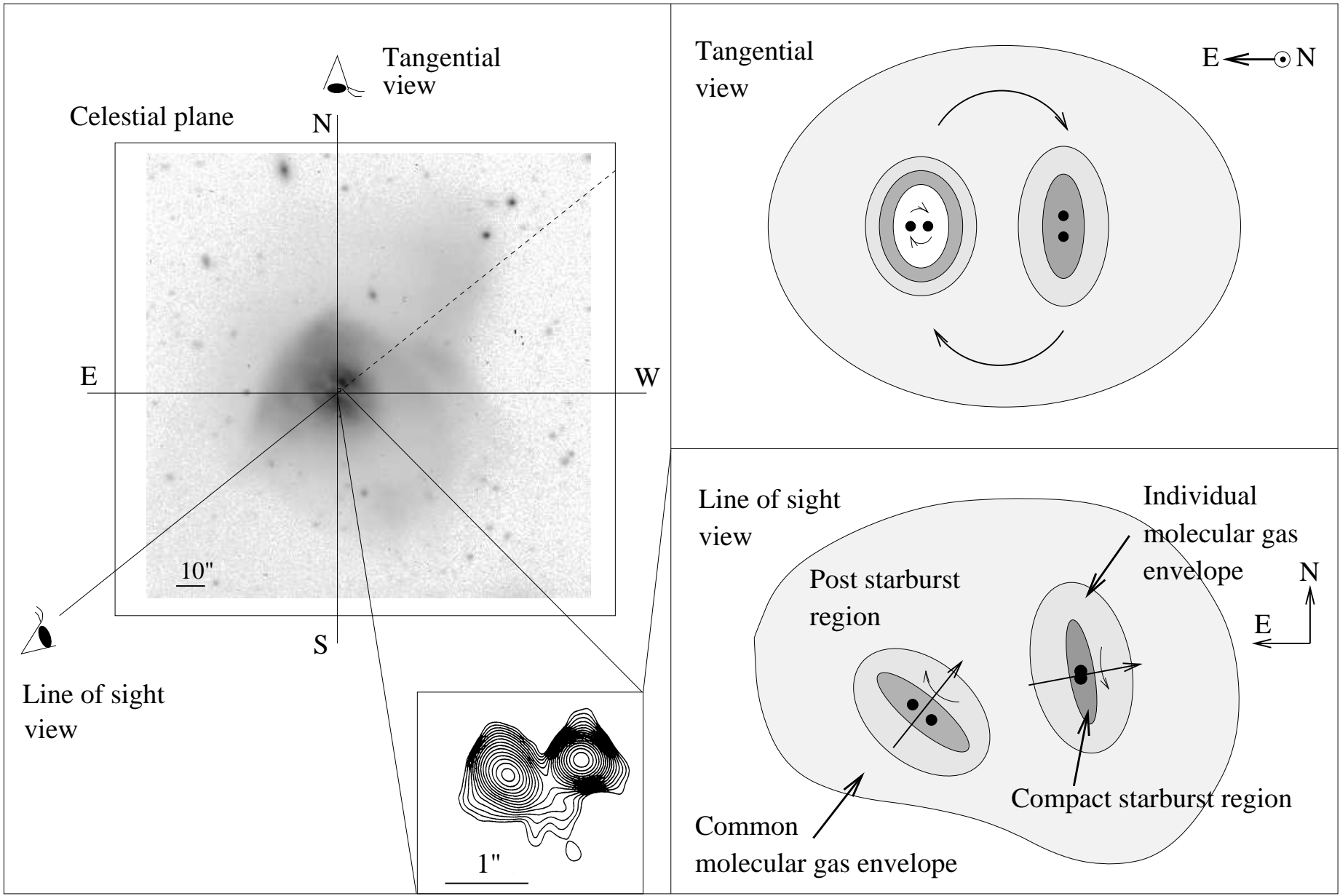
\end{document}